\documentclass[10pt]{article}
\usepackage[OE]{express}
\usepackage{float}

\begin{document}

\title{Simulating $Z_{2}$ topological insulators via a one-dimensional cavity optomechanical cells array}

\author{Lu Qi,\authormark{} Yan Xing,\authormark{} Hong-Fu Wang,\authormark{*} Ai-Dong Zhu,\authormark{} and Shou Zhang\authormark{}}

\address{\authormark{}Department of Physics, College of Science, Yanbian University, Yanji, Jilin 133002, People's Republic of China\\}

\email{\authormark{*}hfwang@ybu.edu.cn} 



\begin{abstract}
We propose a novel scheme to simulate $Z_{2}$ topological insulators via one-dimensional (1D) cavity optomechanical cells array. The direct mapping between 1D cavity optomechanical cells array and 2D quantum spin Hall (QSH) system can be achieved by using diagonalization and dimensional reduction methods. We show that the topological features of the present model can be captured using a 1D generalized Harper equation with an additional $\mathrm{SU}(2)$ guage structure. Interestingly, spin pumping of effective photon-phonon bosons can be naturally derived after scanning the additional periodic parameter, which means that we can realize the transition between different QSH edge states.
\end{abstract}



\section{Introduction}
Topological insulators have become a very active area of research and raised widespread attention since the discovery of quantum Hall (QH) effects~\cite{MC821001,XS831102} in 1980. Normally, the topological insulators exhibit as insulator in the bulk and possess gapless edge states in surface or boundary. These surface states or boundary states are very robust against the disorder and perturbation, as long as the energy gaps of the system remain open, the topologically protected edge states cannot vanish. Besides, the topological classification and the essentially topological features of the system can be captured by a topological invariant. These novel properties stimulate that many theoretical models and experimental schemes have been proposed to investigate topological insulators, including graphene ribbons~\cite{PDG841103,LMWCD23151031},  cold atoms trapped in optical lattices~\cite{YRKJI4620904,FMNEM031305,KLRMAI1081206,FSZCG851207,AWYC1715072}, open system~\cite{HMXX9215081,ZHWX9316082,HWX0414083}, \emph{p}-orbit optical ladder systems~\cite{XEW041308}, off-diagonal bichromatic optical lattices~\cite{SKS801309}, and circuit-QED lattices~\cite{KALS8210101,WZZSMC8612102,BDFFA8713103,FSZCG851210}. It has been verified that these systems can be used to simulate and investigate topological insulators exhibiting QH edge states via the synthetic gauge field, dimensional reduction, etc. methods. The QSH effect, moreover, which is found in materials displaying strong spin-orbit coupling, has opened another new area to investigate $Z_{2}$ topological insulators, a new family of topological states. The intrinsic spin-orbit coupling, as the origin of the QSH effect, determines that the QSH system contains two spin-$\frac{1}{2}$ fermions where the two-component spin fermions can be described as QH states at equal but opposite ``magnetic fields''. Interestingly, this kind of topological insulator state possesses two pairs of edge states with opposite spin components at the boundary of the system. Recently, a number of models and proposals have been put forward to mimic $Z_{2}$ topological insulators. In \cite{NIPAMMI1051011}, a specific 2D tight-binding model has been introduced to simulate a $Z_{2}$ topological insulator by engineering a synthetic gauge field to subject cold atoms. Afterward, Mei $et~al.$~\cite{FSZCG851212} proposed a scheme to simulate $Z_{2}$ topological insulators using dimension reduction method with cold atoms trapped in a 1D optical lattice.

In recent years, with the fast-developing fields of micro-nano manufacturing and materials precessing technology, quantum optical platform is widely used to quantum information processing~\cite{GDWHAS4916121,HAS3914122,HAS211123,HAS871124,HAS3771129,HASXK1911212}, quantum computation~\cite{WSHAS241128,HASK1211210,HASK1311211}, and quantum simulation.
Particularly, cavity optomechanical system~\cite{TK3210813,FS020914,IK030915,MSKN271016} is being one of the most appealing and promising candidates for the study of fundamental quantum physics, including red-sideband laser cooling in the resolved or unresolved sideband regime~\cite{INWT990722,YYXQC911523,ME498224,YKWY901425,FJAS990726,ZTM831127,JFRJJ051628}, motional sidebands of a nanogram-scale oscillator~\cite{MDDHLAKSJ921557}, normal-mode splitting~\cite{JIT1010831,SKMM4600932}, energy transiting~\cite{HDLJ5371655}, coherent-state transiting between the cavity and mechanical resonator~\cite{L841129,YA1081230}, optomechanical buckling transitions~\cite{HUJSJJ081756}, quantum network~\cite{ZWLL911533}, backaction-evading measurements~\cite{AFK100834}, induced transparency~\cite{QJJMZ231540,WYCH061641}, entanglement between mechanical resonator and cavity field or atom~\cite{CDHAS061125,DSAHPAVAM980735,CDP770836,WYYS571438}, macroscopic quantum superposition~\cite{JL1161642}, squeezing light~\cite{KCKMEP790943,TPRNC031344,AFA161445} and squeezing resonator~\cite{DCHAS061127,AJ1030946,WG2113462,FSZCG851247,HGP871348,MGMPGD891449}. Besides, the cavity optomechanical system has the widely promising applications in the field of high-precision measurement, such as micro-mass measurement, micro-displacement measurement, weak force measurement, gravitational-wave detection~\cite{BR218017}, and so on.

In our knowledge, although both cavity optomechanical system and topological insulators have rapidly developed, the simulation of topological features for the topological insulators based on the optomechanical system is rarely investigated yet. Motivated by this, we propose a conceptually simple and theoretically feasible method to simulate $Z_{2}$ topological insulators with a 1D cavity optomechanical cells array. We demonstrate that the original cavity optomechanical array model can be decoupled to two independent bosonic chains, which can be used to simulate $Z_{2}$ topological insulators by introducing a periodic modulation appropriately. The topological features of our model can be described by a generalized 1D Harper equation, and we find that the system exhibits trivial and nontrivial topological phases when the system parameters are chosen to different values. Particularly, the analogous spin pumping of effective photon-phonon bosons can be spontaneously derived with the periodic parameter varying at the range of $(0, 2\pi)$.

The paper is organized as follows. In Section \uppercase\expandafter{\romannumeral2}, we derive the effective Hamiltonian of 1D cavity optomechanical arrays and realize the mapping between our model and 2D QSH system. In Section \uppercase\expandafter{\romannumeral3}, two examples are presented to simulate trivial and nontrivial topological insulators as the system parameters take different forms. We also discuss the $Z_{2}$ topological index which is used to distinguish different QSH phases. And we discuss the experimental feasibility of our scheme before conclusion. A conclusion is given in Section
\uppercase\expandafter{\romannumeral4}.

\section{Model and mapping}
\begin{figure}
  \includegraphics[width=5.5in]{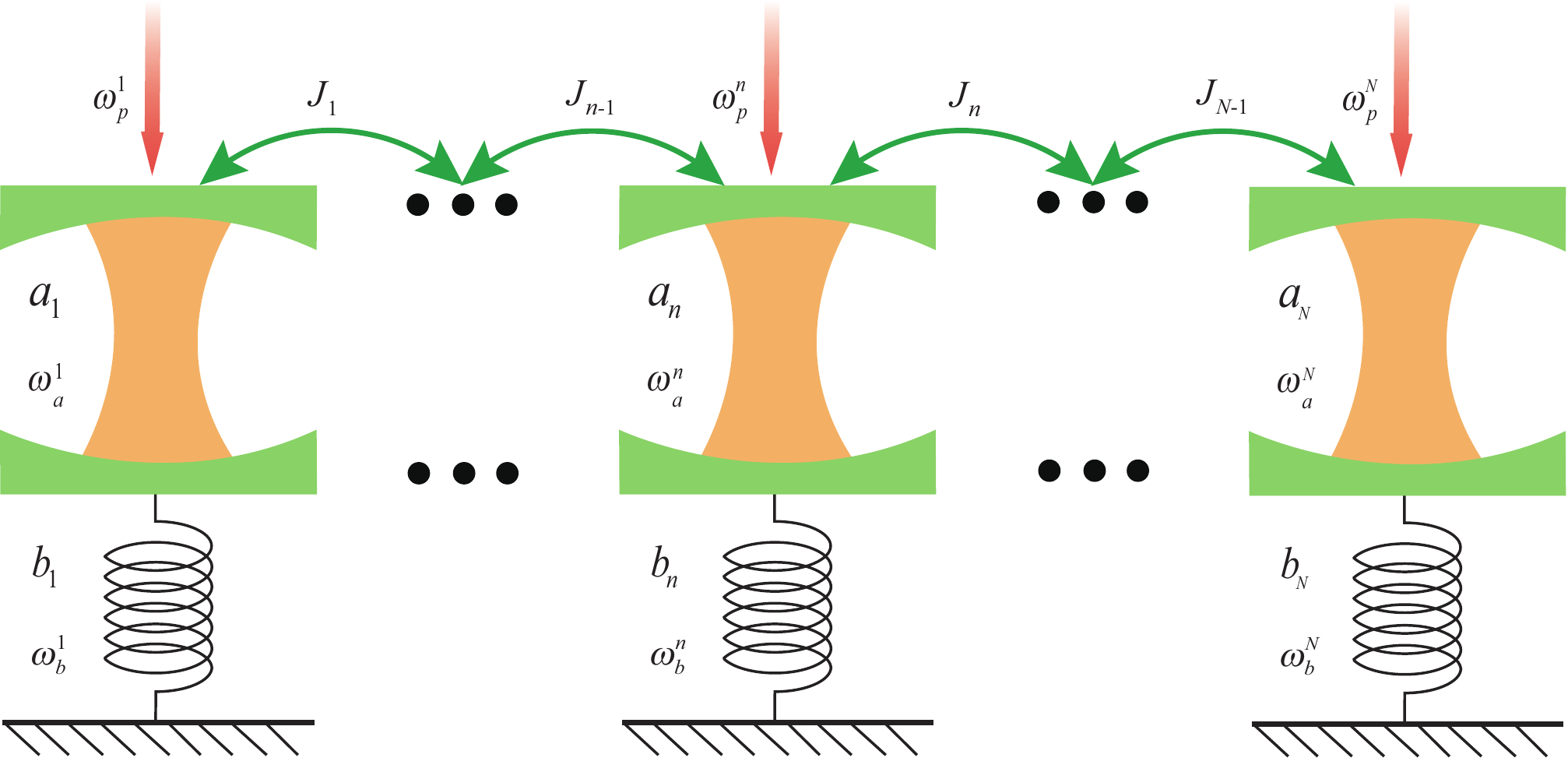}\\
  \caption{Schematic of the 1D cavity optomechanical cells array. The $n$th unit cell contains a cavity mode with frequency $\omega_{a}^{n}$ labelled by $a_{n}$ and a oscillator with frequency $\omega_{b}^{n}$ labelled by $b_{n}$, each cavity mode is driven by a classical pumping field with frequency $\omega_{p}^{n}$. The coupling strengths between the adjacent cavities can be modulated to an appropriate value $J_n$.}\label{fig1}
\end{figure}

The system of 1D cavity optomechanical array we study here as shown in Fig.~\ref{fig1}. In this array, each unit cell is composed of a mechanical resonator with angular frequency $\omega_{b}^{n}$ and a cavity with angular frequency $\omega_{a}^{n}$. The mechanical resonator couples to the cavity which is driven by an external optical field with angular frequency $\omega_{p}^{n}$ and rabi frequency $\Omega_{n}$ via radiation pressure, forming a standard optomechanical subsystem. The total Hamiltonian of the system can be expressed as
\begin{eqnarray}\label{e01}
H^{(1)}&=&\sum_{n}\bigg[\omega_{a}^{n}a_{n}^{\dag}a_{n}+\omega_{b}^{n}b_{n}^{\dag}b_{n}-g_{n}a_{n}^{\dag}a_{n}\left(b_{n}^{\dag}+b_{n}\right)
\cr&&+\Omega_{n}\left(a_{n}e^{i\omega_{p}^{n}t}+a_{n}^{\dag}e^{-i\omega_{p}^{n}t}\right)
+J_{n}\left(a_{n}^{\dag}a_{n+1}+a_{n+1}^{\dag}a_{n}\right)\bigg],
\end{eqnarray}
where $a_{n}^{\dag}$ $(b_{n}^{\dag})$ is the creation operator of the optical cavity (mechanical resonator) for the $n$th unit cell while $a_{n}$ $(b_{n})$ is its corresponding annihilation operator, and $g_{n}$ is the single-phonon optomechanical coupling strength resulted from radiation pressure. The first two terms represent the free energy of the $n$th unit cell, the third term describes the interaction between the cavity and mechanical resonator in $n$th optomechanical unit cell, the forth term stands for the interaction between the cavity field and the external classical laser field, and the last term expresses the hopping between the adjacent optomechanical unit cells with the hopping strength $J_{n}$.

In the rotating frame with respect to the driving frequency $\omega_{p}^{n}$, the above Hamiltonian becomes
\begin{eqnarray}\label{e02}
H^{(2)}&=&\sum_{n}\bigg[-\Delta_{n}a_{n}^{\dag}a_{n}+\omega_{b}^{n}b_{n}^{\dag}b_{n}-g_{n}a_{n}^{\dag}a_{n}\left(b_{n}^{\dag}+b_{n}\right)
\cr&&+\Omega_{n}\left(a_{n}+a_{n}^{\dag}\right)
+J_{n}\left(a_{n}^{\dag}a_{n+1}+a_{n+1}^{\dag}a_{n}\right)\bigg],
\end{eqnarray}
where $-\Delta_{n}=\omega_{a}^{n}-\omega_{p}^{n}$ is the detuning between the cavity field frequency and pumping laser pulse. Under the condition of strong laser driving, we rewrite the operators as the sum of mean values and small quantum fluctuations to linearize Eq.~(\ref{e02}), which means $a_{n}=\alpha_{n}+\delta a_{n}$, $b_{n}=\beta_{n}+\delta b_{n}$. Substituting the above formulas into Eq.~(\ref{e02}), and ignoring the first and third order terms, and further dropping the notation ``$\delta$'' for all the fluctuation operators for the sake of simplicity, the standard linearization Hamiltonian can be obtained as
\begin{eqnarray}\label{e03}
H_{L}&=&\sum_{n}\bigg[-\Delta_{n}a_{n}^{\dag}a_{n}+\omega_{b}^{n}b_{n}^{\dag}b_{n}
-G_{n}\left(a_{n}^{\dag}+a_{n}\right)\left(b_{n}^{\dag}+b_{n}\right)
+J_{n}\left(a_{n}^{\dag}a_{n+1}+a_{n+1}^{\dag}a_{n}\right)\bigg],
\end{eqnarray}
where $G_{n}=g_{n}\alpha_{n}$ is the effective optomechanical coupling strength. For Hamiltonian in Eq.~(\ref{e03}), the 1D cavity optomechanical array can be easily decoupled to two irrelevant bosonic chains by proceeding the diagonalization using the usual Bogoliubov transformation~\cite{HMA620051}.

\begin{figure}
  \includegraphics[width=5.5in]{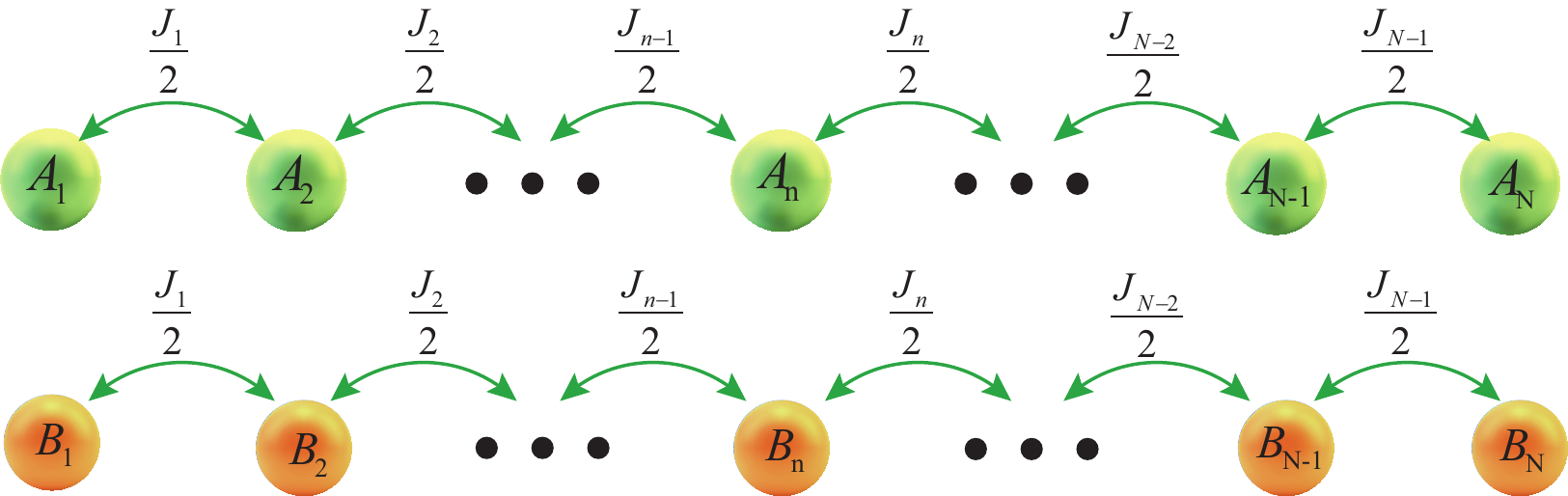}\\
  \caption{Two decoupled bosonic chains. The initial optomechanical array system can be equivalent to two effective irrelevant bosonic chains via diagonalization under the condition of red-detuned regime.}\label{fig2}
\end{figure}

To derive the type of the beam splitter Hamiltonian, we consider the case of red-detuned regime, namely $-\Delta_{n}\approx\omega_{b}^{n}$, in which the Hamiltonian in Eq.~(\ref{e03}) becomes
\begin{eqnarray}\label{e04}
H_{L}^{red}&=&\sum_{n}\bigg[-\Delta_{n}\left(a_{n}^{\dag}a_{n}+b_{n}^{\dag}b_{n}\right)
-G_{n}\left(b_{n}^{\dag}a_{n}+a_{n}^{\dag}b_{n}\right)
+J_{n}\left(a_{n}^{\dag}a_{n+1}+a_{n+1}^{\dag}a_{n}\right)\bigg].
\end{eqnarray}
To obtain the diagonal form of the above Hamiltonian, we make the bose mode transformation as follows
\begin{eqnarray}\label{e05}
{A}_{n}=\frac{(a_{n}+b_{n})}{\sqrt{2}},~~~~~~~~~{B}_{n}=\frac{(a_{n}-b_{n})}{\sqrt{2}}.
\end{eqnarray}
For the case of strongly off-resonant regime $G_{n}\gg J_{n}$, substituting Eq.~(\ref{e05}) into Eq.~(\ref{e04}) and together with the rotating wave approximation, the effective diagonalization Hamiltonian can be written as
\begin{eqnarray}\label{e06}
H&=&\sum_{n}\bigg[\omega_{A}^{n}A_{n}^{\dag}A_{n}+\omega_{B}^{n}B_{n}^{\dag}B_{n}
+\frac{J_{n}}{2}\left(A_{n}^{\dag}A_{n+1}+B_{n}^{\dag}B_{n+1}+\mathrm{H.c.}\right)\bigg],
\end{eqnarray}
where $\omega_{A}^{n}=-\Delta_{n}-G_{n}$ and $\omega_{B}^{n}=-\Delta_{n}+G_{n}$ are the eigenenergy of the effective photon-phonon bosonic modes $A_{n}$ and $B_{n}$, respectively. Equation~(\ref{e06}) shows that the original full Hamiltonian of the 1D cavity optomechanical array can be equivalent to a Hamiltonian consisting of two decoupled bosonic chains, as shown in Fig~\ref{fig2}. The first two terms and last four terms of Eq.~(\ref{e06}) represent the on-site eigenenergy and adjacent hopping interaction of the two bosonic chains respectively. Obviously, the form of Eq.~(\ref{e06}) is consistent with the tight-binding Hamiltonian that is investigated in topological system. It is natural to think that our present 1D optomechanical model can be used to simulate a system that exhibits nontrivial topological phase.

It is worth highlighting that, different from a Fermi system with spin-$\frac{1}{2}$ electron, our model is a standard Bose system, which contains two kinds of bosonic photon and phonon simultaneously. However, the quantized phenomenon in topological insulators originates from its topological properties, and is independent on Fermi statistical distribution of the electron with semi-odd spin. Besides, from the points of energy spectrum and Bloch theorem, electron, photon, and phonon present certain similarity. So, in the field of bosons with integral spin, the quantized topological states of light, sound~\cite{CVOF301758}, mechanical motion and so on should also be implemented similarly.
If we regard the two decoupled bosonic chains as a two-component atomic gas mentioned in \cite{FSZCG851212}, our system with two kinds of effective photon-phonon bosons can be naturally used to simulate the $Z_{2}$ topological insulator with two-compont fermions, in which one of two effective photon-phonon bosonic chains $A_{n}$ can be equivalent to spin-up atomic gas while another can be equivalent to spin-down atomic gas.

In order to simulate the 2D tight-binding model~\cite{NIPAMMI1051011} via the 1D cavity optomechanical cells array, it is significantly necessary to introduce another periodic parameter to map the second dimension of 2D model in present 1D cavity optomechanical cells array. Thus choosing the system parameters $\omega_{A}^{n}$, $\omega_{B}^{n}$, and $\frac{J_{n}}{2}$ as $\omega_{A}^{n}=\lambda\cos(2\pi\beta n+\phi)$, $\omega_{B}^{n}=\lambda\cos(2\pi\beta n-\phi)$, and $\frac{J_{n}}{2}=t$. Here, parameter $\lambda$ is the strength of periodic modulation term introduced into the system, $\beta$ is a real parameter, $\phi$ is an additional phase which can be changed continuously from $0$ to $2\pi$, and $t$ represents the hopping strength. Noticing that this can be achieved by modulating system detuning with respect to driving field $\Delta_{n}=-\lambda\cos(2\pi\beta n)\cos(\phi)$ and effective coupling strength $G_{n}=\lambda\sin(2\pi\beta n)\sin(\phi)$. It has been verified that the manipulation of optomechanical coupling parameter can be relized by acting periodicly modulated driving light or microwave field on the optomechanical system. We can achieve the periodic modulation of system by changing the frequency of driving field in the way of periodic control~\cite{AJ1030946}. Substituting the above periodic parameters into Eq.~(\ref{e06}), the Hamiltonian becomes
\begin{eqnarray}\label{e07}
H&=&\sum_{n}\bigg[\lambda\cos(2\pi\beta n+\phi)A_{n}^{\dag}A_{n}
+\lambda\cos(2\pi\beta n-\phi)B_{n}^{\dag}B_{n}\cr
&&+t\left(A_{n}^{\dag}A_{n+1}+B_{n}^{\dag}B_{n+1}+\mathrm{H.c.}\right)\bigg].
\end{eqnarray}
Obviously, if we associate the operators $A_{n}^{\dag}~(A_{n})$ with the creation (annihilation) operators of the spin-up component fermions (represented by $\uparrow$) and $B_{n}^{\dag}~(B_{n})$ with the creation (annihilation) operators of the spin-down component (represented by $\downarrow$) fermions, regard the parameter $\beta$ as the magnetic flux, and treat the additional phase $\phi$ as the quasimomentum $k_{y}$, our 1D cavity optomechanical cells array can be directly mapped into the 2D setup as in \cite{NIPAMMI1051011} to realize $Z_{2}$ topological insulators. In this configuration, the system can be described by a generalized 1D Harper equation with an additional SU(2) gauge structure as
\begin{eqnarray}\label{e08}
E\Psi_{n,\uparrow\downarrow}(\phi)=M(n,\phi)\Psi_{n,\uparrow\downarrow}(\phi)
+t[\Psi_{n+1,\uparrow\downarrow}(\phi)+\Psi_{n-1,\uparrow\downarrow}(\phi)],
\end{eqnarray}
where
\begin{eqnarray}\label{e09}
M(n,\phi)= \left(
             \begin{array}{cc}
               \lambda\cos(2\pi\beta n+\phi) & 0 \\
               0 & \lambda\cos(2\pi\beta n-\phi) \\
             \end{array}
           \right).
\end{eqnarray}
Based on this equation, the topological phases of our 1D cavity optomechanical array can be captured via the energy spectrum and the distribution of the edge states.

\section{Topological phases}
Considering that we have used the condition $G_{n}\gg J_{n}$ at the moment of decoupling initial Hamiltonian, corresponding to that $\lambda\sin(2\pi\beta n)\sin(\phi) \gg 2t$. Therefore, the effective strength of the periodic modulation must satisfy $\lambda \gg 2\sqrt{2}t$, which is valid as revealed in \cite{YFHYC101654}. In the following we will investigate the spectral properties of Eq.~(\ref{e08}).

\subsection{Trivial topological phase}
\begin{figure}[H]
  \includegraphics[width=5.5in]{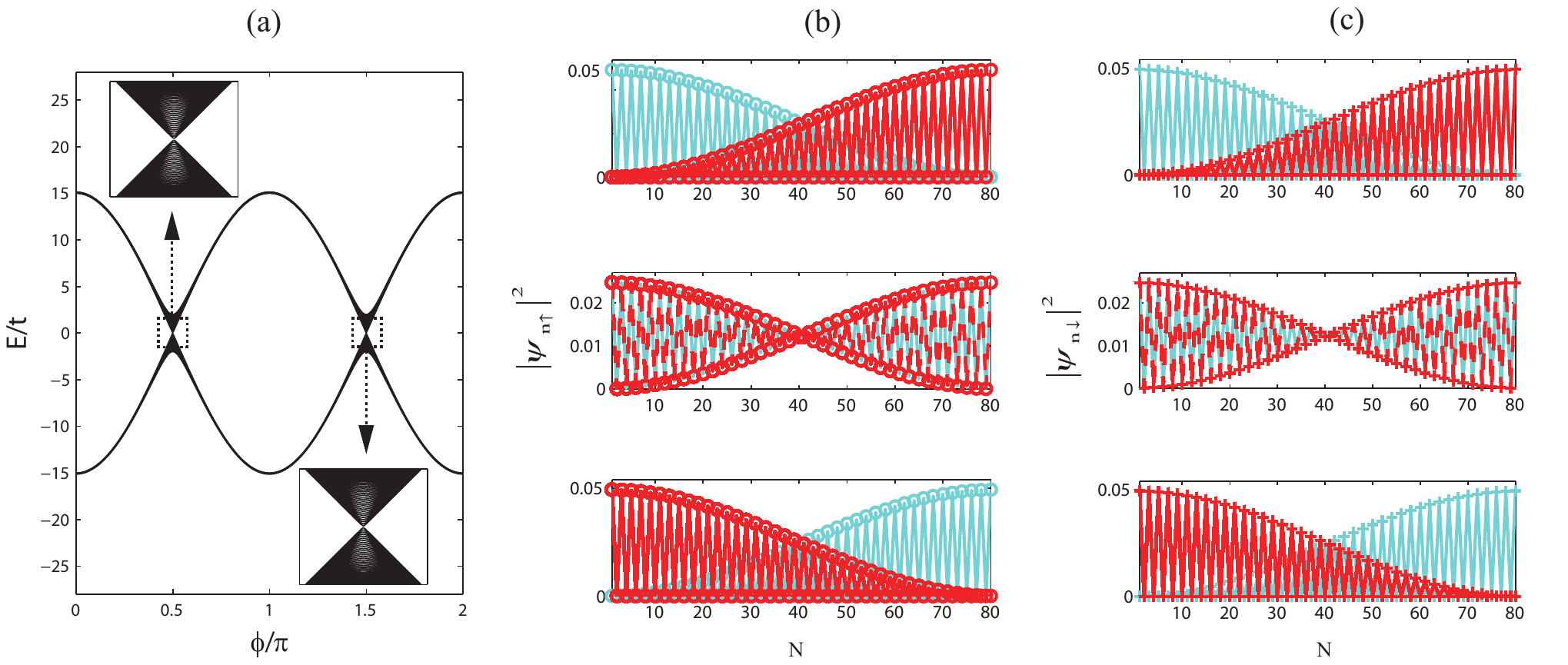}\\
  \caption{Energy spectrum and density distribution. (a) Energy varys with $\phi$. The insets show two Dirac points with a liner dispersion. (b) Distribution of states corresponding to spin up component represented by symbol ``$\circ$''. The up panel corresponding to $\phi=0$, the middle panel corresponding to $\phi=0.5\pi$, and the bottom corresponding to $\phi=\pi$. (b) Distribution of states corresponding to spin down component represented by symbol ``$+$''. The up panel corresponding to $\phi=0$, the middle panel corresponding to $\phi=0.5\pi$, and the bottom corresponding to $\phi=\pi$. The system parameters are chosen as $\beta=\frac{1}{2}t$, $\lambda=15t$, $t=1$, and the lattice size $N=80$.}\label{fig3}
\end{figure}

Firstly, we set the parameters as $\beta=\frac{1}{2}t$, $\lambda=15t$, $t=1$, and $\phi\in(0, 2\pi)$. In this case, each spin component, which is simulated by an effective photon-phonon bosonic chain ($A_{n}$ or $B_{n}$), can be mapped onto a 2D Hofstader model with $\pi$ flux in each plaquette in Eq.~(\ref{e08}), which implies that the system is invariant under the time-reversal transformation for the magnetic flux term only well-defined modulo $2\pi$. Therefore, the system exhibits no QH effect and the QH edge states don't emerge. To demonstrate this point, we plot the energy spectrum of the system under the open boundary condition, as shown in Fig.~\ref{fig3}.

Figure~\ref{fig3}(a) clearly shows that the energy spectrum exists no band gap but contains two Dirac points with a liner dispersion connecting the valence band and conduction band. These results are exactly consistent with the viewpoints mentioned above. To further demonstrate this point, the distributions of wave functions around the Dirac point $(0.5\pi, 0)$ corresponding to different effective photon-phonon bosonic ``spin'' modes ($A_{n}$ and $B_{n}$) are also plotted varying with the number of sites in Figs.~\ref{fig3}(b) and ~\ref{fig3}(c). The results indicate that no matter at the Dirac point or at other place in Brillouin zone, the distributions of wave functions are extended, which shows that there exists no local QH edge modes.

Furthermore, we find that the distributions of the states are completely same in Figs.~\ref{fig3}(b) and ~\ref{fig3}(c), which means that the motor directions of the two ``spin'' components $A_{n}$ and $B_{n}$ are from left to right simultaneously (or from right to left) with $\phi$ varying from $0$ to $2\pi$. The reason is that the matrix in Eq.~(\ref{e09}) becomes $M(n,\phi)=\mathrm{Diag}[\mp\lambda\cos(\phi), \mp\lambda\cos(-\phi)]$ when $\beta$ takes $\frac{1}{2}t$. The periodical modulation term of ``spin-up'' component $A_{n}$ becomes identical with the ``spin-down'' component $B_{n}$ due to the parity of cosine function, which implies that two irrelevant chains degenerate one chain and the system is a topologically trivial $Z_{2}$ insulator.

\subsection{Nontrivial topological phase}
\begin{figure}[H]
  \includegraphics[width=5.0in]{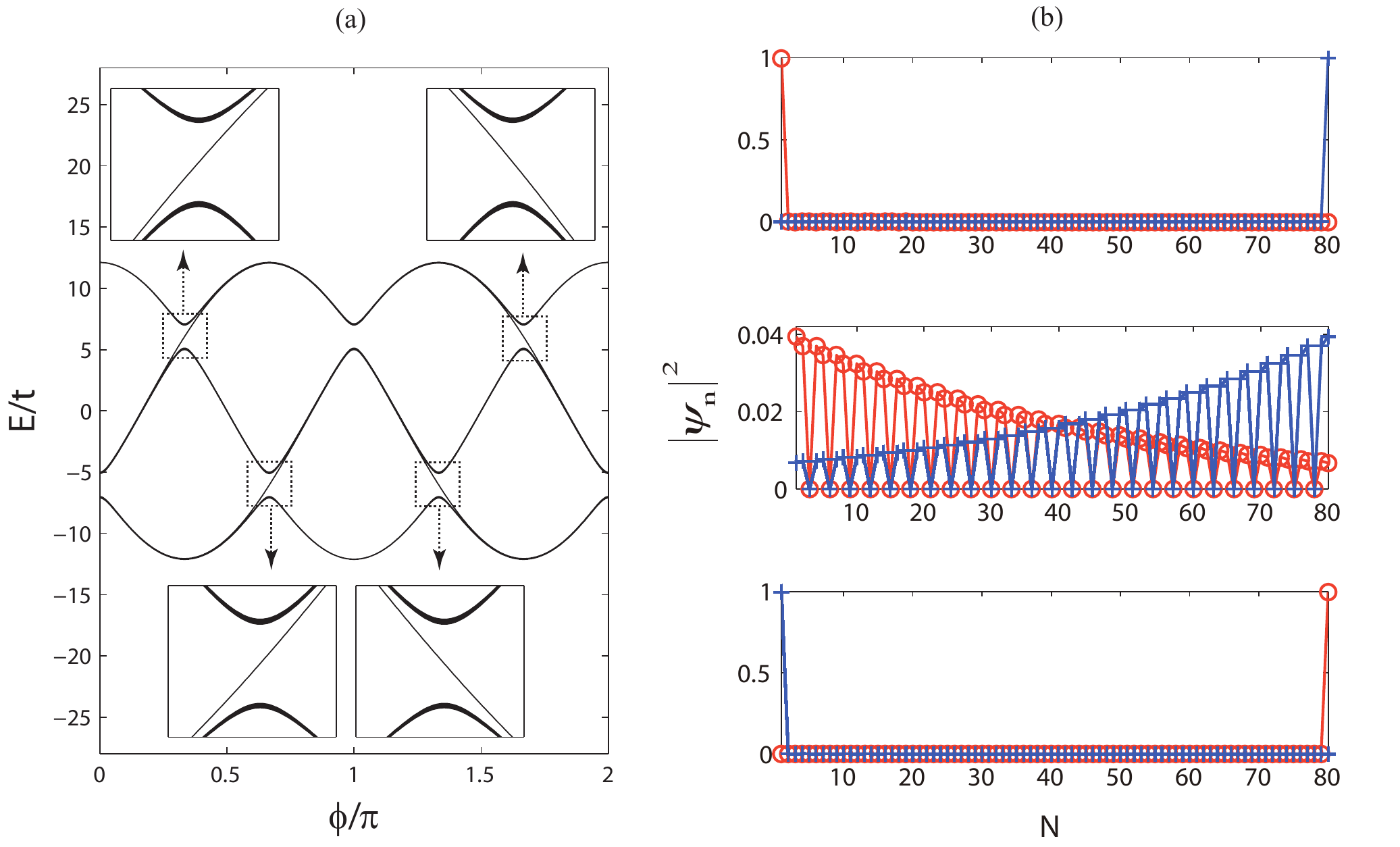}\\
  \caption{Energy spectrum and density distributions. (a) Energy varys with $\phi$. The insets exhibit fine structures corresponding two energy gaps. (b) Density distributions of spin up component represented by symbol ``$\circ$'' and spin down component represented by symbol ``$+$'' vary with $\phi$. The up panel corresponding to $\phi=0.5\pi$, the middle panel corresponding to $\phi=\pi$, and the bottom corresponding to $\phi=1.5\pi$. The system parameters are chosen as $\beta=\frac{1}{3}$, $\lambda=15t$, $t=1$, and the lattice size $N=80$.}\label{fig4}
\end{figure}

Since the case of $\beta=\frac{1}{2}t$ cannot generate the topologically nontrivial phase, we consider $\beta=\frac{1}{3}t$ and keep $\lambda=15t$, $t=1$. Interestingly, we find that the energy spectrum splits into three bulk subbands and appears two band gaps, as shown in Fig.~\ref{fig4}(a). One can see that there exists two degenerate states with opposite ``spins'' that are localized at two ends of the system at each gap, as shown in Fig.~\ref{fig4}(b) (corresponding to the first energy gap). In more detail, for the ``spin-up'' component $A_{n}$, there are two states which localized at the sites $n=1$ and $n=N$ respectively corresponding to $\phi=0.5\pi$ and $\phi=1.5\pi$, as depicted in Fig.~\ref{fig4}(b). Similarly, for the case of ``spin-down'', we find that a state is localized at $n=N$ when $\phi=0.5\pi$ and a state is localized at $n=1$ when $\phi=1.5\pi$, which is just contrary to the ``spin-up'' case. These results indicate that each end of the system exists two degenerate edge states with opposite spins. All these features indicate that our system exhibits QSH effect in analogy to the 2D system which possesses two pairs of helical edge states.

Besides, Fig.~\ref{fig4}(b) also depicts that the ``spin-up'' localized edge state which is at site $1$ will transit to site $N$ (the ``spin-down'' emerges the transition from $N$ to $1$) by adiabatically changing the parameter $\phi$ from $0$ to $2\pi$. We stress that these states are still localized at each edge during the whole process except for the case of $\phi=\pi$ in which the edge states integrate into the bulk state. In the analogous the QSH effect, we find that the above interesting process is the so-called spin pumping. More specifically, the topologically protected edge states with the opposite ``spin'' components can be pumped through the cavity optomechanical arrays by scanning the pumping laser phase adiabatically. This process is an effective way to realize the transition between different topologically protected edge states, which is significant to quantum information processing.

\subsection{Topological index}
As discussed above, if we regard the parameter $\beta$ as the magnetic flux, and treat the additional phase $\phi$ as the quasimomentum $k_{y}$, the 1D cavity optomechanical arrays can be directly mapped into a 2D QSH system. Therefore, the topological properties of our systems can also be characterized by a topological invariant, namely $Z_{2}$ index. When the $Z_{2}$ index of one special system is $0$ and the Femi energy is chosen into the energy gap, it means that the system emerges no QSH edge states and thus it is a trivial system. Similarly, if $Z_{2}$ index of the system is $1$, it means that there exists odd number of edge state pairs at each end of the system and the system is a QSH system. Besides, the index can be guaranteed by the energy gap, which indicates that the $Z_{2}$ index remains constant as long as the energy gap is not closed.

In \cite{FSZCG851212}, Mei $et~al.$ have illuminated that $Z_{2}$ index of a system with decoupled spins can be simply related to the spin Chern number $\nu=\mathrm{SChNmod} 2$, where $\mathrm{SChNmod} 2=(\mathrm{ChN}_{\uparrow}-\mathrm{ChN}_{\downarrow})/2$ and the Chern number of two decoupled spin components $\mathrm{ChN}_{\uparrow(\downarrow)}$ can be individually calculated by
\begin{eqnarray}\label{e10}
\mathrm{ChN}_{\uparrow(\downarrow)}=\frac{1}{2 \pi i}\sum_{E_{\mu} \leq E_{\mathrm{Femi}}}\int dk_{x}d\phi~F[|\Psi_{\uparrow(\downarrow)\mu}(k_{x},\phi)\rangle],
\end{eqnarray}
where $F$ is the Berry curvature with respect to the single particle state $|\Psi_{\uparrow(\downarrow)\mu}(k_{x},\phi)\rangle$, $\mu$ is the energy band index, and the symbol $\sum$ represents the sum of all energy bands below the Femi energy.
Taking this method, we calculate the Chern number of two energy gaps with respect to ``spin-up'' and ``spin-down'' and find that the Chern number corresponding to the first gap is $\mathrm{ChN}_{\uparrow,\downarrow}=\pm1$ and the Chern number corresponding to the second gap is $\mathrm{ChN}_{\uparrow,\downarrow}=\mp1$. Therefore, the $Z_{2}$ index can be derived as $\nu=1$ in both cases, which means that there exists one pair of edge states localized at each end of the system. These results are exactly consistent with the describtion shown in Fig.~\ref{fig3}(b).

\begin{figure}
\centering
  \includegraphics[width=4.0in]{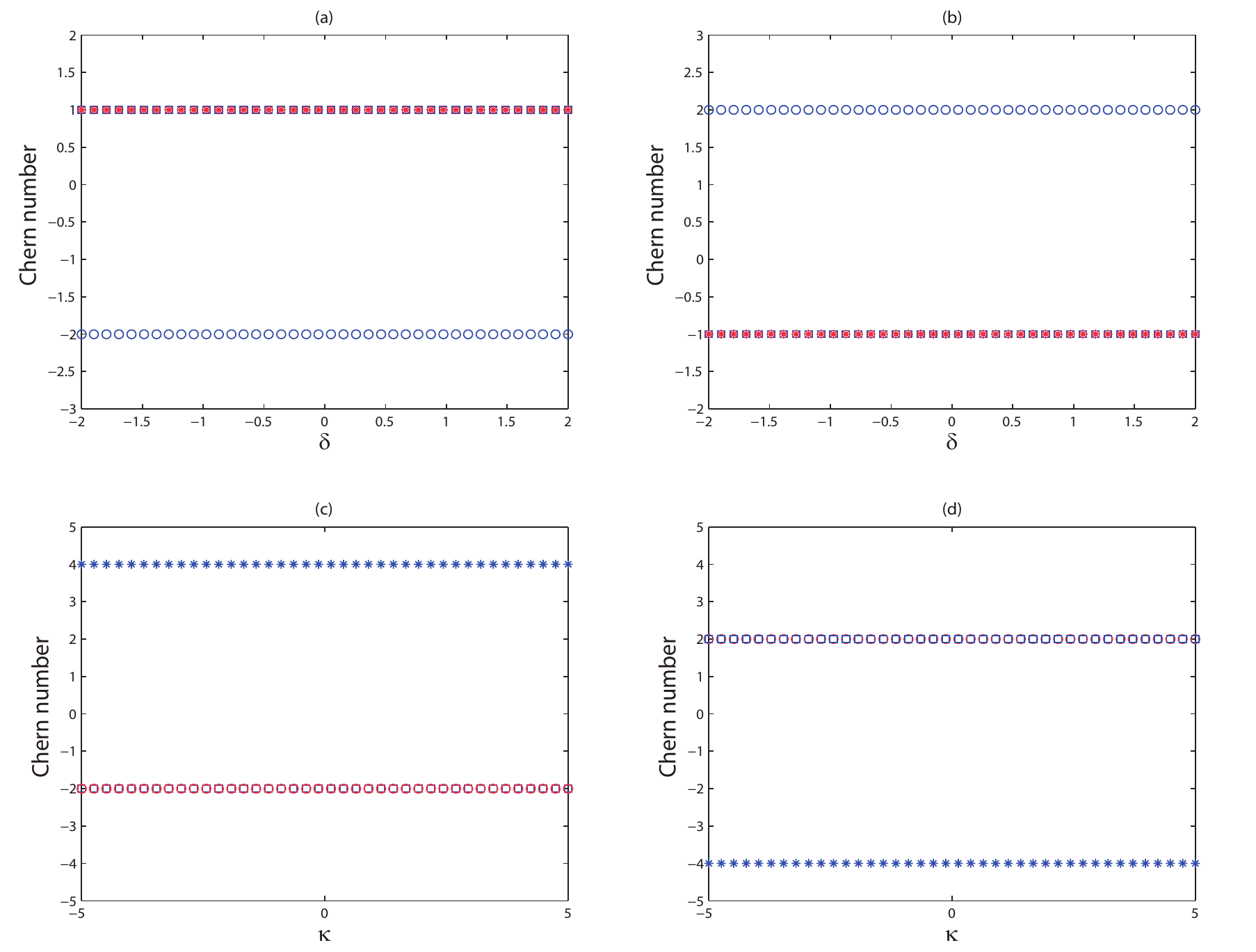}\\
  \caption{(a) and (b) Chern numbers of three energy bands versus the parameter fluctuation $\delta$ corresponding to spin-up component and spin-down component. The square, circle, and red star represent the first, second, and third bands, respectively. (c) and (d) Chern numbers of three energy bands versus the cavity decay $\kappa$ corresponding to spin-up component and spin-down component.  The square, red circle, and star represent the first, second, and third bands, respectively.}\label{fig5}
\end{figure}

As well known, the topological edge states and invariant of the system are very robust against the disorder and perturbation, as long as the energy gaps of the system remain open, the topologically protected edge states cannot vanish. All of the above discussions are under ideal situation, however, the real optomechanical system has parameter fluctuation $\delta$, optical loss $\kappa$, and mechanical noise $\gamma$. In order to investigate the impact of these factors on the system, we numerically calculate the Chern numbers of three energy bands corresponding to two spin components under different cases in Figs.~\ref{fig5} (a) -~\ref{fig5} (d). Figures.~\ref{fig5} (a) and ~\ref{fig5} (b) reveal that a certain parameter range fluctuation of the hopping strength $t$ cannot affect the topology of the system due to the protections of energy gap.
However, when the cavity decay and oscillator damping are under consideration, some new things will occur. We plot the Chern numbers of three energy bands versus the cavity decay (condition $\kappa=\gamma$ is chosen to decouple the Hamiltonian) corresponding to different spin components in Figs.~\ref{fig5} (c) -~\ref{fig5} (d). The results reveal that the Chern numbers change from $1,-2,1$ and $-1,2,-1$  to $-2,-2,4$ and $2,2,-4$ corresponding to spin-up and spin-down components, respectively. Obviously, new topologically nontrivial phases corresponding to different spin components occur due to the introduction of the dissipation. We stress that each spin component is topologically nontrivial, however, the whole system is topologically trivial because of the trivial $Z_{2}$ invariant $Z_{2}=2$ in both cases, which is due to time reversal symmetry is broken by the cavity decay and oscillator damping.

Before conclusion, we now give a brief analysis and discussion in relation to the experimental parameters proposed in our model system. The condition $G_{n}\gg J_{n}$ in our scheme must be satisfied. This condition can be achieved in optomechanical crystals mentioned in \cite{JTAJASMO4781152}, in which the experimental parameters can reach the $\mathrm{GHz}$ regime. In optomechanical crystals, the frequency of oscillators $\omega_{b}^{n}$ and effective optomechanical coupling strength $G_{n}$ can approach $10^{9} \mathrm{Hz}$ order, and the hopping strength between adjacent cavities $J_{n}$ can be tuned $10^{6} \mathrm{Hz}$ order. In this parameters regime, an appropriate example of parameters selection is provided in \cite{GFVS931650}. Even in the work reported in \cite{GMJBF620053}, the effective coupling strength between the cavity and oscillator can reach the range of $\mathrm{THz}$, which means that one can easily tune the parameters to meet the $G_{n}\gg J_{n}$ within current experimentally accessible regimes to check our work.

\section{Conclusions}
In conclusion, we have proposed a novel scheme to simulate a $Z_{2}$ topological insulator based on the 1D optomechanical cells array. We demonstrate that the model can be decoupled to two irrelevant bosonic chains using diagonalization process under the condition of red-detuned regime. Based on the method of dimensional reduction, our model can be directly mapped into a 2D system that exhibits QSH states. Thus we can describe our 1D model by using a spin-$\frac{1}{2}$ generalized Haper equation. Two special examples are given to simulate trivial topological and nontrivial topological $Z_{2}$ topological insulators. We find that the system only exhibits two Dirac points with a linear dispersion when the parameter is chosen as $\beta=\frac{1}{2}$, and the system will emerge two pairs of edge states with opposite spins localized at two ends of 1D arrays when the parameter is chosen as $\beta=\frac{1}{3}$. Moreover, an interesting transition between two opposite nontrivial edges with two spin components can be derived by varying the phase $\phi$ from $0$ to $2\pi$ continuously. Furthermore, we also calculate the topological $Z_{2}$ index to classify the different topological phases of our 1D optomechanical arrays.

\section*{Funding}
National Natural Science Foundation of China under Grant (11465020, 11264042, 61465013, 11564041); The Project of Jilin Science and Technology Development for Leading Talent of Science and Technology Innovation in Middle and Young and Team Project under Grant (20160519022JH).

\section*{Acknowledgments}
We gratefully thank Dr. H. Z. Shen for helpful discussions. 

\end{document}